\documentclass{lmcs}

\usepackage{enumerate}
\usepackage[colorlinks=true]{hyperref}
\usepackage{tikz}
\usepackage{amssymb}

\usepackage{xcolor,latexsym,amsmath,extarrows,alltt}
\usepackage{xspace}
\usepackage{booktabs}
\usepackage{mathtools}
\usepackage{enumitem}
\usepackage{stmaryrd}
\usepackage{microtype}
\usepackage{wrapfig}

\theoremstyle{theorem}\newtheorem{theorem}{Theorem}
\theoremstyle{theorem}\newtheorem{lemma}[theorem]{Lemma}
\theoremstyle{theorem}
\theoremstyle{definition}\newtheorem{definition}[theorem]{Definition}
\theoremstyle{definition}

\newcommand{\B}{\mathcal{B}}

\newcommand{\F}{\mathcal{F}}
\newcommand{\G}{\mathcal{G}}

\newcommand{\V}{\mathcal{V}}

\newcommand{\Terms}{\mathcal{T}}
\newcommand{\Rules}{\mathcal{R}}
\newcommand{\Qules}{\mathcal{Q}}
\newcommand{\Data}{\mathcal{D}\!\!\mathcal{A}}

\newcommand{\arrtype}{\Rightarrow}

\newcommand{\arrz}{\to}
\newcommand{\arr}[1]{\to_{#1}}
\newcommand{\arrr}[1]{\arr{#1}^*}

\newcommand{\symb}[1]{\mathtt{#1}}
\newcommand{\nf}[1]{#1\!\downarrow}
\newcommand{\bitencode}[1]{\overline{#1}}
\newcommand{\termencode}[1]{[#1]}
\newcommand{\trsencode}[1]{\llbracket #1 \rrbracket}
\newcommand{\unknown}[1]{\underline{\symb{#1}}}
\newcommand{\app}[2]{#1 \cdot #2}

\newcommand{\strue}{\symb{true}}
\newcommand{\sfalse}{\symb{false}}
\newcommand{\nul}{\symb{0}}
\newcommand{\one}{\symb{1}}
\newcommand{\nilbits}{\symb{\rhd}}
\newcommand{\Var}{\symb{Var}}
\newcommand{\Fun}{\symb{Fun}}
\newcommand{\Rule}{\symb{Rule}}
\newcommand{\nillist}{\symb{[]}}
\newcommand{\cons}{\symb{::}}
\newcommand{\nilrules}{\symb{\varnothing}}
\newcommand{\bits}{\symb{bitstring}}
\newcommand{\term}{\symb{term}}
\newcommand{\termlist}{\symb{termlist}}
\newcommand{\rules}{\symb{rules}}

\newtheorem{proposition}{Proposition}

\begin{document}

%

\title{Higher-order Cons-free Interpreters}
\thanks{The authors are supported by the Marie Sk{\l}odowska-Curie
action ``HORIP'', program H2020-MSCA-IF-2014, 658162 and by the Danish
Council for Independent Research Sapere Aude grant ``Complexity via
Logic and Algebra'' (COLA).}
\author[C.~Kop]{Cynthia Kop}
\address{Department of Computer Science, Copenhagen University}
\email{kop@di.ku.dk}
\author[J.G.~Simonsen]{Jakob Grue Simonsen}
\address{Department of Computer Science, Copenhagen University}
\email{simonsen@di.ku.dk}

\maketitle

\begin{abstract}
Constructor rewriting systems are said to be cons-free if any constructor term occurring in the rhs
of a rule must be a subterm of the lhs of the rule. Roughly, such systems cannot build new data structures
during their evaluation. In earlier work by several authors, (typed) cons-free systems have been used to characterise
complexity classes such as polynomial or exponential time or space by varying the type orders, and the recursion forms allowed.
This paper concerns the construction of interpreters for cons-free term rewriting.  Due to their connection with proofs by diagonalisation, interpreters may be of use when studying separation results between complexity classes 
in implicit computational complexity theory. We are interested in interpreters of type order $k > 1$ that can interpret any term of strictly lower type order; while this gives us a well-known separation result $\textrm{E}^k\textrm{TIME} \subsetneq \textrm{E}^{k+1}\textrm{TIME}$, the hope is that more refined 
interpreters with syntactically limited constraints can be used to obtain a notion of \emph{faux} diagonalisation and
be used to attack open problems in complexity theory.
\end{abstract}

\section{Introduction}

In~\cite{jon:01}, Jones introduced \emph{cons-free
programming}; roughly \emph{read-only} programs where data structures cannot be created or
altered, only read from the input. 
For example, cons-free programs with data order $0$ can
compute exactly those decision problems which are in PTIME, while
tail-recursive cons-free programs with data order $1$ characterise
$\textrm{PSPACE}$. The field of research studying such characterisations is called
implicit computational complexity (ICC).

Jones' results can easily be generalised to the area of
(higher-order) term rewriting. In term rewriting, systems have no fixed evaluation order (so call-by-name or
call-by-value can be introduced as needed, but are not required),
and reduction is natively non-deterministic.
ICC using cons-free term rewriting has been studied by several authors \cite{car:sim:14,kop:sim:16}.

One important goal of ICC is to provide new separation results between well-known classes. A standard
technique for doing so is \emph{diagonalisation} via interpreters. Roughly, an interpreter for a rewriting system $A$ is a term in another rewriting system $B$ that, when applied to (a bit string representation of) any term from $A$, will simulate evaluation of that term.
It is tantalising to consider interpreters for, and written in, cons-free term rewriting.
While interpreters are very well-known in functional programming, they are very rarely seen in rewriting. Furthermore,
typical programming of self-interpreters in the wild involves maintaining several intermediate data structures, which is patently impossible in cons-free rewriting where
no data constructors are present.

Our work concerns the construction of cons-free interpreters for higher-order term rewriting. As a proof of concept, we consider in this paper an interpreter of type order $2$ that will evaluate any, suitably encoded, term of type order $1$. In future work, we hope
that studying further constraints to the syntactic form of the rules of higher-order systems (effectively, constraining the types of recursion used) may lead to more refined diagonalisation and that such ``faux'' diagonalisation may lead to
results separating known complexity classes.

\section{Preliminaries}

We consider higher-order term rewriting with simple types, and
$\beta$-reduction as a separate step; we reason modulo
$\alpha$-conversion only.  Function symbols are assigned a type
declaration of the form $[\sigma_1 \times \dots \times \sigma_n]
\arrtype \tau$, where $\tau$ does not need to be a base type.
Rules are assumed to have the form $f(\ell_1,\dots,\ell_n) \arrz r$
(so can have a functional type).  we additionally limit interest to
\emph{higher-order constructor TRSs}, which is to say that each
$\ell_i$ in the rule above is a constructor term, and does not
contain either applications or abstractions.

We will use ``data terms'' to refer to the set of ground constructor
terms containing neither abstractions nor applications.
We denote $\Terms(\F,\V)$ for the set of terms built from symbols in
$\F$ and variables in $\V$, and $\Data$ for the set of data terms.
We will particularly consider an \emph{innermost weak} reduction
strategy, which disallows reductions below an abstraction, and
allows a subterm $f(s_1,\dots,s_n)$ to be reduced only if all $s_i$
are abstractions or normal forms.

\begin{definition}
Let $C$ be a class of HOTRSs $(\F,\Rules)$ and $I,O$ be sets indexed
by $C$ (shortly denoted $I_\F$ and $O_\F$ instead of $I_{(\F,\Rules)}$
and $O_{(\F,\Rules)}$) such that $I_\F,O_\F \subseteq \Terms(\F,\V)$.
A HOTRS $(\G,\Qules)$ with start symbol $\symb{simulate} \in \G$ is
an \emph{interpreter} for $C$ \emph{with input set} $I$ and \emph{output set} $O$, if there exist computable
injective functions $\termencode{\cdot}_\F : I_\F  \cup O_\F \mapsto
\Terms(\G,\emptyset)$ and $\trsencode{\cdot} : C \mapsto
\Terms(\G,\emptyset)$ such that, for all $(\F,\Rules) \in
C$ and $a \in I_\F,b \in O_\F$:
\begin{center}
$a \arrr{\Rules} b$ if and only if
$\symb{simulate}(\trsencode{(\F,\Rules)},\termencode{a}_\F) \arrr{
\Qules} \termencode{b}_\F$
\end{center}
\end{definition}

\section{Cons-free Term Rewriting}

Like Jones~\cite{jon:01}, we limit interest to \emph{cons-free}
rules, adapted to term rewriting as follows:

\begin{definition}[Cons-free Rules]
A rule $\ell \arrz r$, presented using $\alpha$-conversion in a form 
where all binders are distinct from the free variables, is
\emph{cons-free} if for all subterms $s = f(s_1,\dots,s_n)$ of $r$
with $f$ a constructor, either $s$ is a subterm of $\ell$ or, if not,
$s$ is a data term.
A left-linear (higher-order) constructor TRS $(\F,\Rules)$ is
cons-free if all rules in $\Rules$ are. 
\end{definition}

Cons-free term rewriting enjoys many convenient properties.  Most
importantly, the set of data terms that may be reduced to using
cons-free rules is limited by the data terms in the start term and
the right-hand sides of rules, as described by the following
definition:

\begin{definition}
For a given ground term $s$, the set $\B_s$ contains all data terms
$t$ which occur as (a) a subterm of $s$ or (b) a subterm of the
right-hand side of some rule in $\Rules$.
\end{definition}

$\B_s$ is a set of data terms, is closed under subterms and, since
$\Rules$ is fixed, has a linear number of elements in the size of $s$.
The property that no new data is generated by reducing $s$ is
formally expressed by the following result:

\begin{definition}[$\B$-safety]
Let $\B \subseteq \Data$ be a set which (a) is closed under subterms,
and (b) contains all data terms occurring in the right-hand side of a
rule in $\Rules$.  A term $s$ is $\B$-safe if for all subterms $t$ of
$s$: if $t = c(t_1,\dots,t_n)$ with $c$ a constructor, then $t \in
\B$.
\end{definition}

\begin{lemma}\label{lem:safetypreserve}
Let $\Rules$ be cons-free.
For all $s,t$:
if $s$ is $\B$-safe and $s \arrr{\Rules} t$, then $t$ is $\B$-safe.
\end{lemma}

Thus, for a decision problem $\symb{start}(s_1,\dots,s_n)
\arrr{\Rules} t$ (where $t$ and all $s_i$ are data terms), all terms
occurring in the reduction are $\B$-safe.  This insight allows us to 
limit interest to $\B$-safe terms in most cases, and is instrumental
to obtain the following results:

\begin{proposition}\label{prop:exptime}
The class of decision problems in $\mathsf{EXP}^k\mathsf{TIME}$ contains
exactly those functions which can be accepted by:
\begin{itemize}
\item a cons-free HOTRS of order $k$ with a weak-innermost reduction
  strategy;
\item a cons-free \emph{confluent} HOTRS of order $k$ with a
  weak-innermost reduction strategy.
\end{itemize}
\end{proposition}

That is, adding determinism does not make a difference to the
characterisation result.  Hence for simplicity we will focus on
confluent TRSs in particular.

\section{Interpretations}

We consider a system with the following constructors:
\[
\begin{array}{rclcrcl}
\nul & : & [\bits] \arrtype \bits & &
\Var & : & [\bits] \arrtype \term \\
\one & : & [\bits] \arrtype \bits & &
\Fun & : & [\bits \times \termlist] \arrtype \term \\
\nilbits & : & \bits & &
\bot & : & \term \\
\nillist & : & \termlist & &
\nilrules & : & \rules \\
\cons & : & [\term \times \termlist] \arrtype \termlist & &
\Rule & : & [\term \times \term \times \rules] \arrtype \rules \\
\end{array}
\]
Bit strings can be used to encode numbers in the usual way,
e.g.~$\bitencode{6} = \one(\nul(\one(\nul(\nilbits))))$; we assume a
unique encoding for each bitstring, so without leading zeros.
To encode a first-order TRS $(\F,\Rules)$, we enumerate the
function symbols, writing $\F =\{f_1,\dots,f_k\}$, and in each rule
$\rho$ we assume the variables are in $\{x_1,\dots,x_p\}$ for some
$p$.  For the term encoding, let:
\[
\begin{array}{rcl}
\termencode{x_i}_\F & = & \Var(\bitencode{i}) \\
\termencode{f_i(s_1,\dots,s_n)}_\F & = & \Fun(\bitencode{i},
  \termencode{s_1}_\F\ \cons\ \dots\ \cons\ 
  \termencode{s_n}_\F\ \cons\ \nillist) \\
\end{array}
\]
Here, the list constructor $\cons$ is denoted in an infix,
right-associative way. 
Writing $\Rules = \{\ell_1 \arrz r_1,\dots,
\ell_m \arrz r_m\}$, the TRS is encoded as the following term:
\[
\Rule(\termencode{\ell_1},\termencode{r_1},
\Rule(\termencode{\ell_2},\termencode{r_2},\Rule(\dots,
\Rule(\termencode{\ell_m},\termencode{r_m},\nilrules)\dots)))
\]
Of course, strictly speaking an intepreter should operate on bit
strings only.  We have chosen for this more verbose encoding because
it is easier to understand the resulting interpreter-system, and the
same ideas can be transferred to a more restrictive encoding.

\medskip
We seek to define a confluent cons-free HOTRS $(\G,\Qules)$ with
a weak-innermost reduction strategy
which, given a confluent, cons-free first-order TRS $(\F,\Rules)$
with innermost reduction, and a data term $w$, obtains
the encoding for $\nf{w}$, provided this is a data term.  Formally,
if $(\F,\Rules)$ is encoded as the term $R$, we must have
$\symb{normalform}(R,\termencode{w}_\F) \arrr{\Qules}
\termencode{\nf{w}}_\F$ for any ground start term $w$ with $\nf{w}_\F$
a data term, and $\symb{normalform}(R,\termencode{w}_\F) \arrr{\Qules}
\bot$ if $\nf{w}_\F$ is not a data term.  This can be used
to determine whether $w \arrr{\Rules} \strue$ (but is more general).

Note that, since $\Qules$ must be cons-free, we cannot
represent intermediate terms (e.g.~the direct reduct of the start
term).
Instead,
$\Qules$ will operate on \emph{tuples} ($\termencode{s}_\F$,$\gamma$),
where $\gamma$ is a ``substitution'': a term of type
$\bits \arrtype \term$ mapping the representations of variables in
$s$ to data terms or $\bot$ (which indicates any term
in normal form that is not a data term).

\medskip
To work!  We will recurse over the set of rules, but carry along the
complete set, as well as the arguments to the start term (which
together define the set $\B$), for later use.
\[
\begin{array}{rcl}
\symb{normalform}(R,\Fun(f,args)) & \arrz &
  \symb{normalise}(\Fun(f,args),\lambda x.\bot,R,args) \\
\symb{normalise}(\Var(x),\gamma,R,bs) & \arrz & \app{\gamma}{x} \\
\symb{normalise}(\Fun(f,args),\gamma,R,bs) & \arrz &
  \symb{findrule}(\Fun(f,args),\gamma,R,R,bs) \\
\symb{findrule}(w,\gamma,\nilrules,R,bs) & \arrz &
  \symb{substitute}(w,\gamma,bs,R) \\
\symb{findrule}(w,\gamma,\Rule(\ell,r,tl),R,bs) & \arrz &
  \symb{test}(\symb{match}(w,\gamma,\ell,\lambda x.\bot,R,bs),w,
  \gamma,\ell,r,tl,R,bs) \\
\symb{test}(\delta,w,\gamma,\ell,r,tl,R,bs) & \arrz &
  \symb{test2}(\app{\delta}{\nilbits},\delta,w,\gamma,\ell,r,tl,R,bs)
  \\
\symb{test2}(\bot,\delta,w,\gamma,\ell,r,tl,R,bs) & \arrz &
  \symb{normalise}(r,\delta,R,bs) \\
\symb{test2}(\Var(\nilbits),\delta,w,\gamma,\ell,r,tl,R,bs) & \arrz &
  \symb{findrule}(w,\gamma,tl,R,bs) \\
\end{array}
\]
Thus, we normalise $w\gamma$ just by substituting if $w$ is a
variable or no rules match (so $w\gamma$ is in normal form either
way); $\symb{substitute}(w,\gamma,bs,R)$ reduces to $\bot$ if
$w\gamma$ is not a data term.  The function
$\symb{match}$ is used to test whether a rule matches and find the
relevant substitution $\delta$ in one go: in case of a match,
$\app{\delta}{t}$ reduces to $\bot$ for every $t$ which does not
refer to a variable in $\ell$, and in case of no match, it reduces to
$\Var(\nilbits)$ instead (which is not a representation of any term).
In the case of a successful match, we continue to normalise $r\delta$.

To define $\symb{substitute}$, we note that $w$ is always a subterm of
either the start term, or the right-hand side of a rule, and is not a
variable; the result of substituting is a normal form, so we must
reduce to a data term---which, by Lemma~\ref{lem:safetypreserve}, is
a subterm of $bs$ or of the right-hand side of some rule---or to
$\bot$.  These observations give the following rules:
\[
\begin{array}{rcl}
\symb{eqbits}(\nilbits,\nilbits) & \arrz & \strue \\
\symb{eqbits}(\nilbits,\unknown{b}(ys)) & \arrz & \sfalse\ \ 
  \hfill\llbracket\text{for}\ \unknown{b} \in \{\nul,\one\}\rrbracket \\
\symb{eqbits}(\unknown{a}(xs),\nilbits) & \arrz & \sfalse\ \ 
  \hfill\llbracket\text{for}\ \unknown{a} \in \{\nul,\one\}\rrbracket \\
\symb{eqbits}(\unknown{a}(xs),\unknown{a}(ys)) & \arrz &
  \symb{eqbits}(xs,ys)\ \ 
  \hfill\llbracket\text{for}\ \unknown{a} \in \{\nul,\one\}\rrbracket \\
\symb{eqbits}(\unknown{a}(xs),\unknown{b}(ys)) & \arrz & \sfalse\ \ 
  \hfill\llbracket\text{for}\ \unknown{a},\unknown{b} \in \{\nul,\one\}
  \wedge \unknown{a} \neq \unknown{b}\rrbracket \\
\end{array}
\]
\[
\begin{array}{rcl}
\symb{eqsubst}(\Var(x),\gamma,t) & \arrz &
  \symb{eqsubst}(\app{\gamma}{x},\lambda y.\bot,t) \\
\symb{eqsubst}(\bot,\gamma,t) & \arrz & \sfalse \\
\symb{eqsubst}(\Fun(f,as),\gamma,\Var(y)) & \arrz & \sfalse \\
\symb{eqsubst}(\Fun(f,as),\gamma,\Fun(g,bs)) & \arrz &
  \symb{eqcheck}(\symb{eqbits}(f,g),as,\gamma,bs) \\
\symb{eqcheck}(\sfalse,as,\gamma,bs) & \arrz & \sfalse \\
\symb{eqcheck}(\strue,\nillist,\gamma,\nillist) & \arrz & \strue \\
\symb{eqcheck}(\strue,s\cons ss,\gamma,t \cons ts) & \arrz &
  \symb{eqcheck}(\symb{eqsubst}(s,\gamma,t),ss,\gamma,ts) \\
\end{array}
\]
\[
\begin{array}{rcl}
\symb{substitute}(w,\gamma,bs,R) & \arrz &
  \symb{substcheckbs}(\symb{subst}(w,\gamma,bs),w,\gamma,R) \\
\symb{subst}(w,\gamma,\nillist) & \arrz & \bot \\
\symb{subst}(w,\gamma,b\cons bs) & \arrz &
  \symb{subst2}(\symb{eqsubst}(w,\gamma,b),b,w,\gamma,bs) \\
\symb{subst2}(\strue,b,w,\gamma,bs) & \arrz & b \\
\symb{subst2}(\sfalse,\Fun(f,as),w,\gamma,bs) & \arrz &
  \symb{subst3}(\symb{subst}(w,\gamma,as),w,\gamma,bs) \\
\symb{subst3}(\bot,w,\gamma,bs) & \arrz & \symb{subst}(w,\gamma,bs) \\
\symb{subst3}(\Fun(f,as),w,\gamma,bs) & \arrz & \Fun(f,as) \\
\symb{substcheckbs}(\Fun(f,as),w,\gamma,R) & \arrz & \Fun(f,as) \\
\symb{substcheckbs}(\bot,w,\gamma,R) & \arrz & \dots \\
\end{array}
\]
Thus, to obtain $w\gamma$, we find the subterm of the start term or
the rules equal to it.  The rules for the latter case---verifying that
$w\gamma$ is a constructor-term and finding a subterm of the
right-hand side of a rule equal to $w\gamma$, or reducing to $\bot$ if
either part fails---have been omitted for space reasons.

The next task is to find whether $w\gamma$ instantiates some left-hand
side $\ell$, and obtain the relevant substitution if so.  Here,
``instantiates'' should not be taken literally as $w$ is not
necessarily a basic term.  Rather, writing $w = f(w_1,\dots,w_n)$, we
seek to confirm whether $f(\nf{w_1\gamma},\dots,\nf{w_n}\gamma) =
\ell\delta$ for some $\delta$.  Note that $\ell$ is necessarily
left-linear, and its strict subterms are constructor terms.
\[
\begin{array}{rcl}
\symb{match}(\Fun(f,ss),\gamma,\Fun(g,ts),\delta,R,bs) & \arrz &
  \symb{matchcheck}(\symb{eqbits}(f,g),ss,\gamma,ts,\delta,R,bs) \\
\symb{matchcheck}(\sfalse,ss,\gamma,ts,\delta,R,bs) & \arrz &
  \lambda x.\Var(\nilbits) \\
\symb{matchcheck}(\strue,ss,\gamma,ts,\delta,R,bs) & \arrz &
  \symb{matchall}(ss,\gamma,ts,\delta,R,bs) \\
\symb{matchall}(\nillist,\gamma,\nillist,\delta,R,bs) & \arrz &
  \delta \\
\symb{matchall}(s\cons ss,\gamma,t\cons ts,\delta,R,bs) & \arrz &
  \symb{matchall}(ss,\gamma,ts,\symb{instantiate}(\\
  & & \hfill\symb{normalise}(s,\gamma,R,bs),t,\delta),R,bs) \\
\end{array}
\]
\[
\begin{array}{rcl}
\symb{instantiate}(w,\Var(y),\delta) & \arrz &
  \lambda x.\symb{ifelse}(\symb{eqbits}(x,y),w,\app{\delta}{x}) \\
\symb{instantiate}(\bot,\Fun(g,ts),\delta) & \arrz &
  \lambda x.\Var(\nilbits) \\
\symb{instantiate}(\Fun(f,ss),\Fun(g,ts),\delta) & \arrz &
  \symb{install}(\symb{eqbits}(f,g),ss,ts,\delta) \\
\symb{install}(\sfalse,ss,ts,\delta) & \arrz &
  \lambda x.\Var(\nilbits) \\
\symb{install}(\strue,\nillist,\nillist,\delta) & \arrz & \delta \\
\symb{install}(\strue,s\cons ss,t\cons ts,\delta) & \arrz &
  \symb{install}(\strue,ss,tt,\symb{instantiate}(s,t,\delta)) \\
\symb{ifelse}(\strue,x,y) & \arrz & x \\
\symb{ifelse}(\sfalse,x,y) & \arrz & y \\
\end{array}
\]
Thus, $\symb{match}(f(\vec{w}),\gamma,f(\vec{\ell})\delta,R,bs)$
iterates over both $\vec{w}$ and $\vec{\ell}$, updating $\delta$
for each $w_i$ such that $\nf{w_i\gamma} = \ell_i\delta$.  This uses
the $\symb{instantiate}$ helper function, which assumes that $w\gamma$
is a data term or $\bot$.  If any of the instantiations fails (so if
the rule does not match), $\delta$ is updated to have $\app{\delta}{
t} = \Var(\nilbits)$ for all lists $t$ which do not correspond to a
variable in $w$.

\section{Conclusions}

Thus we have seen:

\begin{center}\it
There is a second-order cons-free confluent HOTRS \\which is an
interpreter for the class $C$ of first-order cons-free confluent TRSs.
\end{center}

Assuming Proposition~\ref{prop:exptime} holds, we could now use a
diagonalisation argument to obtain a new proof for the hierarchy
result $\mathsf{PTIME} \neq \mathsf{EXPTIME}$.  Generalising the
program to higher orders, we should also be able to obtain that each
$\mathsf{EXP}^k\mathsf{TIME} \neq \mathsf{EXP}^{k+1}\mathsf{TIME}$.

While these are known results, the ideas used in this proof might
transfer to more ambitious projects.  For example, if we could give a
\emph{tail recursive} variation of the interpreter then, following
Jones~\cite{jon:01}, we might obtain a proof of the proposition
$\mathsf{PTIME} \neq \mathsf{PSPACE}$.

\bibliography{references}
\bibliographystyle{plain}

\end{document}